\documentstyle[psfig]{mn}

\begin{document}
\title{On the streaming motions of haloes and galaxies}
\author[R. K. Sheth, A. Diaferio, L. Hui \& R. Scoccimarro]
{Ravi K. Sheth$^1$, Antonaldo Diaferio$^2$, Lam Hui$^{3,4}$ \&
Rom\'an Scoccimarro$^4$\\
$^1$ NASA/Fermilab Astrophysics Group, MS 209, Batavia, IL 60510-0500\\
$^2$ Dipartimento di Fisica Generale ``Amedeo Avogadro'', 
Universit\`a di Torino, Italy\\
$^3$ Department of Physics, Columbia University, 538 West 120th Street, 
New York, NY 10027\\
$^4$ Institute for Advanced Study, School of Natural Sciences, Einstein Drive, Princeton, NJ 08540\\
\smallskip
Email: sheth@fnal.gov, diaferio@ph.unito.it, scoccima@ias.edu, lhui@ias.edu}
\date{2000 October 5}

\maketitle

\begin{abstract}
A simple model of how objects of different masses stream towards 
each other as they cluster gravitationally is described.  
The model shows how the mean streaming velocity of dark matter 
particles is related to the motions of the parent dark matter haloes.  
It also provides a reasonably accurate description of how the 
pairwise velocity dispersion of dark matter particles differs from that 
of the parent haloes. 
The analysis is then extended to describe the streaming motions of 
galaxies.  This shows explicitly that the streaming motions measured in 
a given galaxy sample depend on how the sample was selected, and shows 
how to account for this dependence on sample selection.  In addition,
we show that the pairwise dispersion should also depend on sample 
type.  Our model predicts that, on small scales, redshift space 
distortions should affect red galaxies more strongly than blue.  
\end{abstract} 


\section{Introduction}\label{intro}
Gravity makes objects cluster.  Therefore, the motions of objects 
towards each other may provide information about the background 
cosmology.  Of course, different subsets of the clustering particles 
may trace the underlying streaming motions differently.  
The scale dependence of the mean streaming $v_{12}^{\rm dm}(r)$ of 
dark matter particles has been understood for some time now 
(Hamilton et al. 1991; Nityananda \& Padmanabhan 1994).  
But there has been little study of how this statistic depends on 
trace-particle type.  

To do this, we build a model in which gravitational clustering is 
viewed as the combination of two processes.  The first arises from the 
fact that gravity causes matter to stream towards local minima of the 
gravitational potential.  This requires a model of how matter which was 
initially distributed rather smoothly around the centre of collapse 
becomes redistributed into a more centrally concentrated density profile 
as the collapse proceeds.  The second process is that these centres 
around which local collapses are occurring, these clusters, are themselves 
moving towards each other:  clusters cluster.  It is the combination of 
these two types of motions which gives rise to the spatial distribution 
and streaming motions of objects today.  

Section~\ref{v12old} summarizes useful results which follow from linear 
theory.  Section~\ref{2halo} shows how the streaming motions of collapsed 
dark matter haloes depend on halo mass.  
Section~\ref{dm} uses this to model the streaming motions of particles, 
rather than haloes.  It shows what fraction of a particle's streaming 
motion arises from the motion of its parent halo, and what fraction 
must arise from motions within the halo.  These smaller scale motions 
are essentially a consequence of the collapse around the halo centre 
we referred to earlier.  It then presents measurements from numerical 
simulations which show that the model predictions are reasonably accurate.  
It also shows that the model provides a reasonable description of how 
the second moment of the pairwise velocity distribution of the dark matter 
differs from that of haloes.  

Section~\ref{gals} shows how to extend the model to study the mean 
streaming motions and the pairwise velocity dispersion of galaxies 
and presents measurements from semianalytic galaxy formation simulations 
which show that the model predictions are reasonably accurate.  
Section~\ref{discuss} discusses what this model implies if one wishes 
to use measurements of the streaming motions of galaxies to make 
inferences about cosmology.  

\section{The model}
\subsection{The mean streaming velocity}\label{v12old}
We will begin by reviewing the strategy which led to the 
derivation of how $v_{12}^{\rm dm}(r)$ depends on scale.  
The relevant starting equation is the pair conservation 
equation in Peebles' book (Peebles 1980), but we will start with 
the equation in the form presented by Nityananda \& Padmanabhan (1994):
\begin{equation}
{\partial\ (1+\bar\xi)\over\partial\,\ln a} = 
-{v_{12}(r)\over Hr}\ 3\,\Bigl[1+\xi(r)\Bigr]
\label{pairs}
\end{equation}
where $\bar\xi(r,a)$ is the volume averaged correlation function 
on proper (rather than comoving) scale $r$ at the 
time when the expansion factor is $a$, and the Hubble constant is $H$.  
This says that if we know the correlation function for all scales $r$ 
and all times $a$, then the assumption that the number of pairs is 
conserved allows us to compute how $v_{12}(r)$ depends on scale today.

An approximate solution to this expression can be got as follows 
(Peebles 1980).  Assume that $\bar\xi$ evolves according to linear 
theory:  $\bar\xi(r,a) = [D(a)/D_0]^2 \bar\xi(r,a_0)$, where $D(a)$ 
is the linear theory growth factor at $a$, and $D_0$ is the growth 
factor at the present time when $a=a_0$.  In an Einstein de-Sitter 
cosmology, $D(a)/D_0 = a/a_0$.  
Then the left hand side is  
$\partial\ \bar\xi(r,a)/\partial\,\ln a = 2\,f(\Omega)\,\bar\xi(r,a)$, 
where $f(\Omega)\equiv \partial\,\ln D/\partial\,\ln a$.  
So, in this approximation we get 
\begin{equation}
-{v_{12}(r)\over Hr} = {2f(\Omega)\over 3}
{\bar\xi(r,a)\over 1 + \xi(r,a)}.
\label{v12lin}
\end{equation}
On large scales, $\xi\ll 1$, and so this is just the usual linear 
theory expression with an extra factor of $(1 + \xi)$ in the 
denominator.  While this approximation is fine on large scales 
($r\ge 10$ Mpc$/h$), it underestimates the exact solution by a factor 
of 3/2 or so on smaller scales (Juszkiewicz, Springel \& Durrer 1998; 
Sheth et al. 2000).  

Hamilton et al. (1991) showed they could compute a good estimate of 
the evolution of $\xi(r,a)$, if the initial correlation function is 
known (also see Nityananda \& Padmanabhan 1994).  
Hamilton et al. also showed that by inserting their expression for 
the evolution of $\xi(r,a)$ into equation~(\ref{pairs}) above, they 
were able to describe the shape of $v_{12}^{\rm dm}(r)$ well on all 
scales.  

While this approach is very useful for studying the statistics of 
dark matter particles, it is not obvious that it can be used to estimate 
the streaming motions of galaxies.  This is because one usually assumes 
that galaxies form at different times.  This means that the number of 
galaxies is not conserved, so the number of galaxy pairs is not conserved.
This means, for example, that the correlation function of galaxies 
refers to different sets of particles at different times.  Therefore,
there is little reason to expect that inserting the correlation 
function of galaxies into the pair conservation equation should 
provide a good estimate of $v_{12}^{\rm gal}(r)$ today.  We show below 
that, provided one makes the correct choice of what one uses for 
$\xi_{\rm gal}(r,a)$, the pair conservation equation can be used to 
provide an accurate estimate of the streaming motions of galaxies.  
 
\subsection{The haloes}\label{2halo}
This subsection is concerned with the first moment of the pairwise 
velocity distribution of haloes identified at the present time.  
Every halo will be represented by one particle, say, the one at 
the halo centre of mass today.  
Imagine tracing these centre-of-mass particles back in time.  
By definition the number of these particles is conserved, 
since all we're doing is following them back to high redshift.  
Of course, at high redshift, few if any of the haloes would actually 
have collapsed around these centre-of-mass particles.  Nevertheless, 
we will use the motions of these particles to represent the motions of 
the halo centre of mass.  Peebles' pair conservation equation, combined 
with the assumption that the motion of a halo today is the same as that 
of its associated centre-of-mass particle, says that if we knew $\xi(r,a)$ 
for these tracer particles, then we can compute $v_{12}^{\rm halo}(r)$ 
today.  

So, to compute $v_{12}^{\rm halo}$, we are stuck with the problem of 
studying the spatial distribution (i.e., the bias factor) of a special 
marked set of particles at earlier times.  The case in which the marked 
particles (in this case, the halo centres-of-mass) are observed at 
a later epoch than when they were marked is familiar: e.g. this is 
like the Mo \& White (1996) simple model for galaxies, in which galaxies 
formed in haloes at $z=3$ but we only observe them today.  
Here, we are interested in the spatial distribution of the special 
particles at earlier epochs than when they were marked.  

The halo centre-of-mass particles are biased tracers of the dark matter 
distribution.  The large scale bias factor is the square root of the 
ratio of the correlation function of these particles to that of the 
dark matter correlation function on large scales.  It depends on halo 
mass:
\begin{equation}
\xi_{\rm hh}(r) \approx b^2(m) \xi^{\rm Lin}_{\rm dm}(r), 
\label{xihhb}
\end{equation}
and a similar equality holds for $\bar\xi_{\rm hh}(r)$.  Here 
$\xi^{\rm Lin}_{\rm dm}(r)$ denotes the initial correlation function 
of the dark matter extrapolated using linear theory to the present time.  
In equation~(\ref{xihhb}) we use the linearly extrapolated 
$\xi^{\rm Lin}_{\rm dm}(r)$ rather than the present day $\xi(r)$ as a 
practical way of taking into account the volume exclusion effects of 
haloes at small scales. 
See Section~2 in Sheth et al. (2000) for a detailed discussion.

To a good approximation, 
\begin{equation}
 b(m) = 1 + {\nu^2(m)-1\over \delta_{\rm c0}\ D(a)/D_0},
 \label{biasm}
\end{equation}
where $\nu(m) \equiv \delta_{\rm c0}/\sigma(m)$ is a function which 
increases with decreasing halo mass, and $D(a)$ and $D_0$ were 
defined earlier.  At the present time, $D(a)=D_0$ and this is the 
familiar Eulerian bias formula from Mo \& White (1996).  
The Lagrangian bias factor is usually expressed as the ratio of 
$\xi_{\rm hh}$ at the initial time to the linearly extrapolated 
$\xi^{\rm Lin}_{\rm dm}$.  This means that the Lagrangian bias factor is 
\begin{displaymath}
 b_{\rm Lag}(m) = \sqrt{\xi_{\rm hh}(r)\over\xi^{\rm Lin}_{\rm dm}(r)}\,
                        {D(a_i)\over D_0}
                  = {D(a_i)\over D_0} + {\nu^2(m)-1\over \delta_{\rm c0}}
\end{displaymath}
where $a_i$ denotes the expansion factor at the initial time.  
Since $a_i\ll a_0$, $b_{\rm Lag}\to (\nu^2 - 1)/\delta_{\rm c0}$,  
which is another familiar expression from Mo \& White (1996).  
So, in this approximation, 
\begin{equation}
 {\partial\ b(m)\over\partial\,\ln a} = f(\Omega)\,\Bigl[1-b(m)\Bigr].
\label{ba}
\end{equation}

\begin{figure*}
\centering
\mbox{\psfig{figure=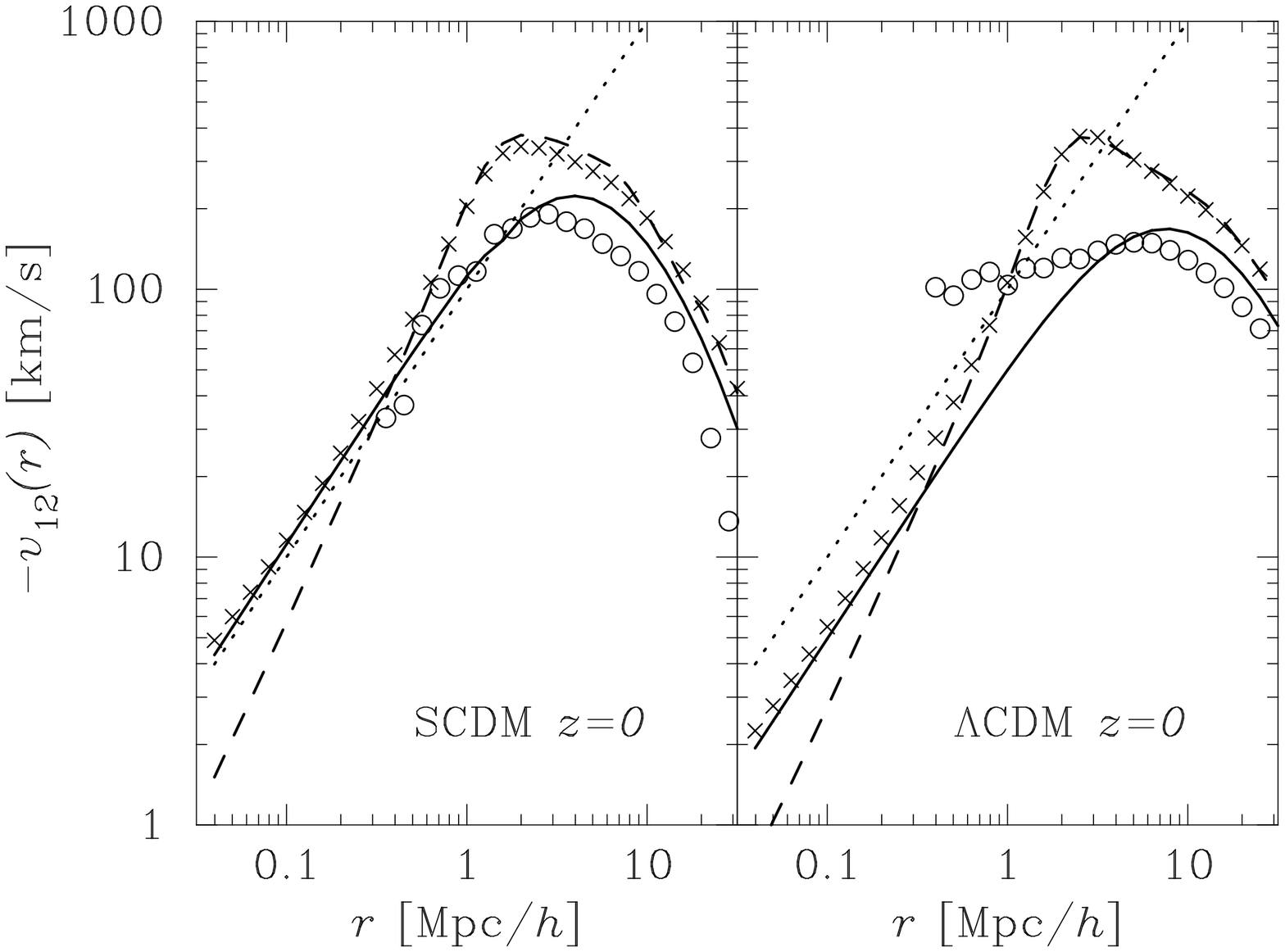,height=6cm,bbllx=68pt,bblly=59pt,bburx=613pt,bbury=459pt}}
\mbox{\psfig{figure=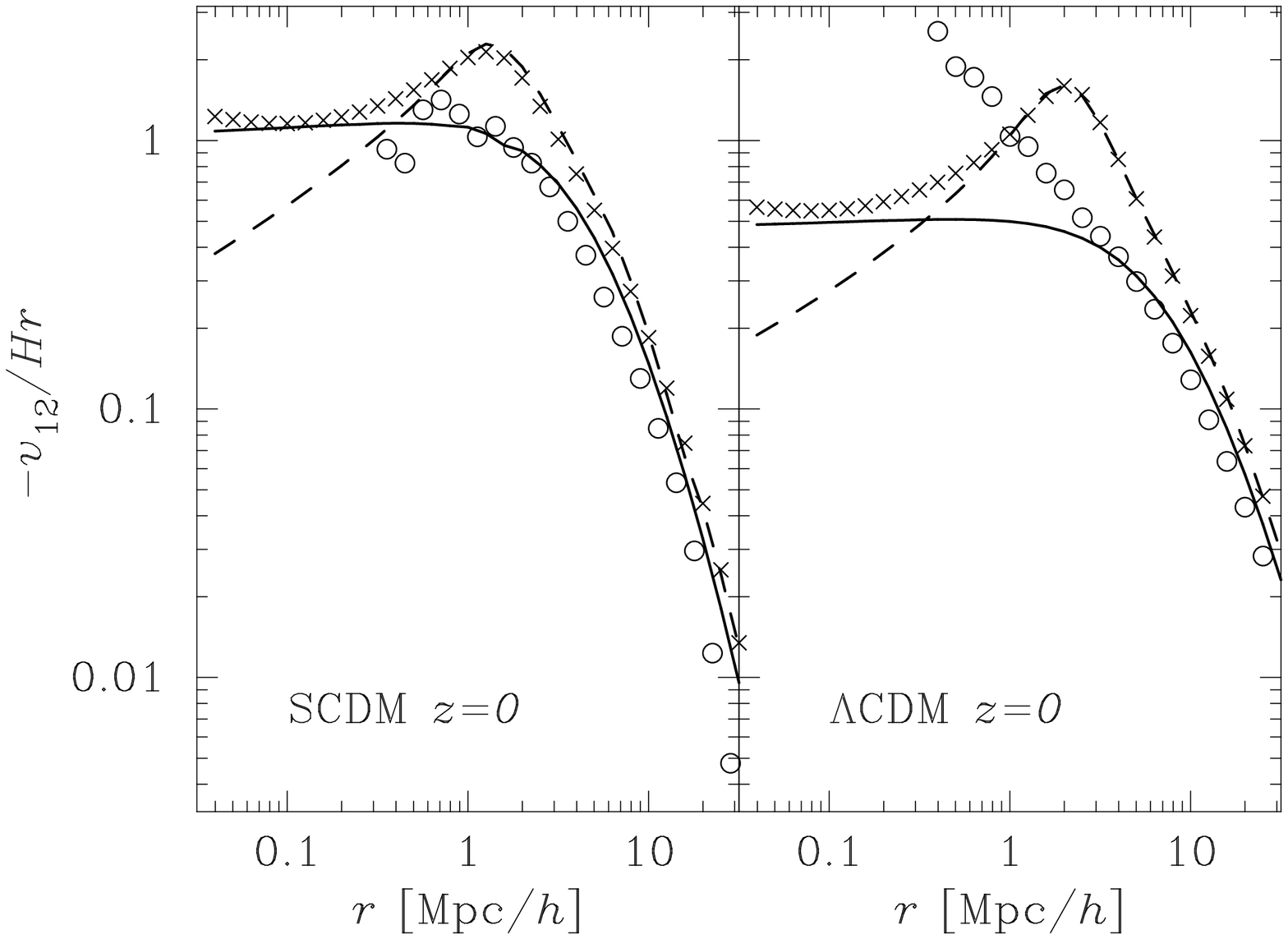,height=6cm,bbllx=69pt,bblly=59pt,bburx=613pt,bbury=452pt}}
\caption{Mean streaming motions in the GIF simulations.  
Open circles and crosses show the streaming motions of the haloes 
and the dark matter, respectively.  Solid and dashed curves show our 
predictions for the haloes (equation~\ref{v12hav}), which should be 
accurate on large scales, and the dark matter 
(the sum of equations~\ref{v12twoh} and~\ref{lamseq}), 
which should be accurate on all scales.  
Dotted lines show the Hubble velocity for comparison.}
\label{gifhaloes}
\end{figure*}

It is straightforward to insert these expressions for the halo correlation 
function and its evolution into the pair conservation formula 
(equation~\ref{pairs}) to see how different $v_{12}^{\rm halo}$ is from 
$v_{12}^{\rm dm}$.  If we study the streaming motions of haloes of two 
different masses, then we must replace $b^2(m)\to b(m_1)\,b(m_2)$.  
This gives  
\begin{eqnarray}
{v_{12}^{\rm halo}(r)\over Hr} &=& {v_{12}^{\rm dm}(r)\over Hr}
{b_1b_2 [1+\xi^{\rm Lin}_{\rm dm}(r)] \over 
        [1 + b_1b_2\,\xi^{\rm Lin}_{\rm dm}(r)]} \nonumber\\
&&\ - {f(\Omega)\bar\xi^{\rm Lin}_{\rm dm}(r)\over 3}
{[b_1(1-b_2) + b_2(1-b_1)]\over[1 + b_1b_2\,\xi^{\rm Lin}_{\rm dm}(r)]}.
\label{v12h}
\end{eqnarray}
If we insert the linear evolution approximation for the relation 
between the correlation function and $v_{12}$ (equation~\ref{v12lin}), 
then this becomes 
\begin{equation}
{v_{12}^{\rm halo}(r)\over Hr} \approx {v_{12}^{\rm dm}(r)\over Hr}
\ \left({b_1 + b_2\over 2}\right)
\ {1+\xi^{\rm Lin}_{\rm dm}(r) \over 1 + b_1b_2\,\xi^{\rm Lin}_{\rm dm}(r)}.  
\label{v12hlin}
\end{equation}
Notice that when $b_1=b_2=1$, then  
 $v_{12}^{\rm halo}(r) = v_{12}^{\rm dm}(r)$.  
Also, in the large separation (small $\bar\xi$) limit,  
$v_{12}^{\rm halo}(r)\to (b_1+b_2)/2$ times $v_{12}^{\rm dm}(r)$.  
So, on average and on large separations, relative to the dark matter, 
massive haloes $(b_1+b_2)>2$ stream towards each other whereas less 
massive haloes $(b_1+b_2)<2$ stream away from each other.  
This makes some physical sense; 
clusters cluster, so they are moving towards each other, whereas smaller 
clumps are in or at the edges of expanding voids, so they are separating 
from each other.  This linear bias of the streaming velocities at large 
separation is consistent with the linear theory analysis of 
Fisher et al. (1994).  

Notice that $v_{12}$ scales with the sum of the bias factors.  
If one ignored the evolution of the bias factor when using 
equation~(\ref{v12lin}), one would have concluded that the scaling 
was with the product of the bias factors---including the evolution of the 
bias factor is essential to getting the correct answer.  Finally, notice 
that on smaller scales where $\xi^{\rm Lin}_{\rm dm}>1$, this analysis 
suggests that $v_{12}$ of less massive haloes should be larger than that 
of the dark matter, with the opposite trend being true for massive haloes.  
Of course, the linear theory and linear evolution approximations we used 
to obtain equation~(\ref{v12hlin}) are not accurate on small scales.  
Nevertheless, this provides at least some indication of the small scale 
behaviour of the halo streaming motions.  

Fig.~\ref{gifhaloes} compares this model with measurements in the 
SCDM and $\Lambda$CDM GIF (Kauffmann et al. 1999) simulations which 
were run by the Virgo collaboration (Jenkins et al. 1999) and are now 
available to the public.  
The open circles show the streaming motions of all the haloes 
with $m>2\times 10^{11}M_\odot/h$ and $m>8.4\times 10^{11}M_\odot/h$ 
in the SCDM and the $\Lambda$CDM simulations, and the solid line shows 
what our model predicts.  Specifically, it shows 
\begin{eqnarray}
-{V_{12}^{\rm halo}(r)\over Hr} &\equiv& 
-\int {\rm d}m_1\int{\rm d}m_2 {v_{12}^{\rm halo}(r)\over Hr} \nonumber \\
&&\ \times \ {n(m_1)n(m_2) [1+b(m_1)b(m_2)\xi^{\rm Lin}_{\rm dm}(r)]\over 
\bar n^2_{\rm halo}[1 + b^2_{\rm halo}\xi^{\rm Lin}_{\rm dm}(r)]}\nonumber \\
&=& {2f(\Omega)\over 3} 
{b_{\rm halo}\bar\xi^{\rm Lin}_{\rm dm}(r)\over 
1 + b^2_{\rm halo}\xi^{\rm Lin}_{\rm dm}(r)}.  
\label{v12hav}
\end{eqnarray}
where $\bar n_{\rm halo}\equiv \int {\rm d}m\,n(m)$ is the average 
number density of haloes,  
$b_{\rm halo}\equiv \int {\rm d}m\,n(m)b(m)$ is their average 
bias factor, the weighting factor in the second line is the ratio of the 
number of $m_1$ and $m_2$ halo pairs at $r$ to the total number 
of halo pairs at $r$, and the final expression follows from inserting 
equation~(\ref{v12hlin}) for $v_{12}^{\rm halo}$ and using 
equation~(\ref{v12lin}) for $v_{12}^{\rm dm}$.  Our model, which we 
only expect to be accurate on large scales because our approximation 
for the halo correlation function, equation~(\ref{xihhb}), breaks down 
on small scales, is reasonably accurate down to scales of order a 
Mpc$/h$ or so.  For comparison, the dotted curves show the Hubble velocity.  
The crosses show the streaming motions of the dark matter particles 
in the simulations, and the dashed curve shows the prediction 
associated with the model described in the next section.  

\subsection{The dark matter}\label{dm}
The large scale net streaming motion of the dark matter can be got from 
our expression for the halo motions by integrating up the contribution 
to the streaming motion from pairs in different mass haloes, weighting 
by the fraction of the total number of pairs which are in such haloes, 
and weighting by the halo mass function:  
\begin{eqnarray}
-{v_{12}^{\rm 2halo}(r)\over Hr} &=&
-\int\! {\rm d}m_2 \int\! {\rm d}m_1\,{m_1 n(m_1)\over\bar\rho}
{m_2 n(m_2)\over\bar\rho} \nonumber \\
&&\qquad\times\qquad {[1 + b(m_1)b(m_2)\xi^{\rm Lin}_{\rm dm}(r)]\over 
1 + \xi_{\rm dm}(r)}\,{v_{12}^{\rm halo}(r)\over Hr}\nonumber\\
&=&-{v_{12}^{\rm dm}(r)\over Hr}\,{1 + \xi^{\rm Lin}_{\rm dm}(r)\over 
1 + \xi_{\rm dm}(r)}.
\label{v12twoh}
\end{eqnarray}
The final equality follows from inserting equation~(\ref{v12h}), 
noting that $\int {\rm d}m\, mn(m)\equiv \bar\rho$, 
and using the fact that the bias factors are defined so that 
$\int {\rm d}m\, mn(m)\,b(m)\equiv \bar\rho$.

We can now make two important points.  
The first is that, at large separations, this expression equals 
$v_{12}^{\rm dm}(r)=v_{12}^{\rm 2halo}(r)$; in this regime the 
streaming motions of the dark matter particles are entirely due to 
the fact that the haloes which contain the particles are moving.  
Moreover, in this regime, $v_{12}^{\rm dm}(r)\approx v_{12}^{\rm L}(r)$, 
where $v_{12}^{\rm L}$ is got from equation~(\ref{v12lin}) by using 
the linear theory values of $\xi$ and $\bar\xi$.
The second is that this expression exactly equals that in 
Sheth et al. (2000) for the contribution to $v_{12}(r)$ from particles 
which are in separate haloes (see their eq.~19).  
This will be important in what follows.  

Notice that on smaller scales, 
 $\xi^{\rm Lin}_{\rm dm}(r) < \xi_{\rm dm}(r)$.  
In this regime the halo motions only account for a fraction of 
$v_{12}^{\rm dm}(r)$.  The remaining contribution to $v_{12}^{\rm dm}(r)$ 
must arise from the streaming motions of pairs in which both particles 
are in the same halo.  This means that the fact that our model for halo 
motions is not accurate on small scales will not matter very much for 
the small scale value of $v_{12}^{\rm dm}$ because, on small scales, the 
fraction of pairs which are in separate haloes, and so are affected by 
this inaccuracy, is small.  We turn, therefore, to a discussion of the 
streaming motions of pairs in which both particles are in the same halo.  

If haloes are stable, then the streaming motion within a halo exactly 
cancels the Hubble flow:  $-v_{12}(r)/Hr = 1$.  In this case, the 
contribution from pairs which are in the stable haloes equals unity 
times 
\begin{displaymath}
\int {\rm d}m {m^2n(m)\over \bar\rho^2}{\lambda(r|m)\over 1+\xi_{\rm dm}(r)}
= {\xi_{\rm 1halo}(r)\over 1 + \xi_{\rm dm}(r)},
\end{displaymath}
where $m^2\lambda(r|m)$ denotes the number of pairs at separation $r$ 
which are in the same halo which has mass $m$; it depends on 
the density profiles of haloes.  For all halo shapes of interest, 
this expression approaches unity at very small $r$, because 
$\xi_{\rm dm}(r) = \xi_{\rm 1halo}(r) + \xi^{\rm Lin}_{\rm dm}(r)\approx 
\xi_{\rm 1halo}(r)$,
and $\xi_{\rm dm}(r)\gg 1$ on scales which are smaller than a typical halo.  
So, if stable clustering is correct, then $-v_{12}^{\rm dm}(r)/Hr = 1$ 
on small scales.  In fact, the mean pairwise velocity on small scales 
depends on the low-mass behaviour of $n(m)$ and $\lambda(r|m)$---in general, 
there is no guarantee that $n(m)$ and $\lambda(r|m)$ will conspire to give 
stable clustering (Ma \& Fry 2000; Sheth et al. 2000).  

In particular, Section~4 of Sheth et al. (2000) shows that the small 
scale term is 
\begin{eqnarray}
-{v^{\rm 1halo}_{12}\over Hr}\! &=&\! {\partial\over\partial{\rm ln}a}
\int {\rm d}m\,{m^2n(m,a)\over\bar\rho^2}\, \int_0^r \!\!
{{\rm d}y\over r}\, {y^2\over r^2}{\lambda(y|m,a)\over[1+\xi(r,a)]}
\nonumber \\
&=& {\partial{\rm ln}\,m_*(a)\over\partial{\rm ln}a}
{[\bar\xi_{\rm 1halo}(r,a) - \xi_{\rm 1halo}(r,a)]\over 3[1 + \xi(r,a)]},
\label{lamseq}
\end{eqnarray}
where $\lambda(r|m,a)$ is proportional to the number of pairs 
in the same $m$-halo which have separation $r$, and $n(m,a)$ is the 
number density of virialized $m$-haloes at time $a$.  
(Our notation differs slightly from that in Sheth et al.---they 
absorbed the two factors of $m$ into their definition of $\lambda$, 
whereas here we have chosen to show these factors explicitly.  Hence, 
our $m^2\lambda$ is their $\lambda$.)
In Sheth et al., $n(m,a)$ depended on time, and the halo profile 
did as well, because virialized haloes were assumed to have profiles 
of the form given by Navarro, Frenk \& White (1997), and these 
halo shapes depend both on $m/m_*(a)$ and on the ratio of the average 
density within the virialized halo to the background density at the 
time it virialized.  (Strictly speaking, the expression above assumes 
that all of the time dependence of the halo shape can be written as a 
function of $m/m_*$---see Sheth et al. for details.)  

The sum of equations~(\ref{v12twoh}) and~(\ref{lamseq}) gives a complete 
description of the streaming motions of dark matter on all scales.  
The dashed curves in Fig.~\ref{gifhaloes} show that this sum provides 
a good description of the dark matter streaming motions on all but 
the smallest scales (see Sheth et al. 2000 for a discussion of the 
discrepancy at the smallest scales).  Now, the dashed curves correspond  
to independently identifying haloes at each epoch. On the other hand, we 
could have followed the approach of Section~2.2---identify or mark the 
haloes today, and then follow them backward in time.  With this latter 
approach, the number density of haloes is fixed to the value today, 
$n(m, a_0)$, but the profile changes from, say a tophat to a more 
centrally concentrated shape. Equation~(\ref{v12twoh}) shows that the 
contribution to the streaming motions from particles in separate haloes 
is independent of the details of how this happens.  
However, recall that this contribution exactly equals the two-halo 
contribution to $v^{\rm dm}_{12}$ worked out by Sheth et al. (2000).  
This means that the one-halo contributions to $v_{12}$ must also be 
the same in both approaches.  
In particular, this means that however the profile changes from a 
tophat to an NFW shape, it must change in just such a way that the 
final answer for the streaming motions of the dark matter particles 
equals equation~(\ref{lamseq}).  Indeed, we can use this requirement 
to constrain how the profile changes from the initial tophat to the 
final NFW cusp---the Appendix shows a worked example of how to do this.

\subsection{The pairwise velocity dispersion}\label{vdisp}

\begin{figure}
\centering
\mbox{\psfig{figure=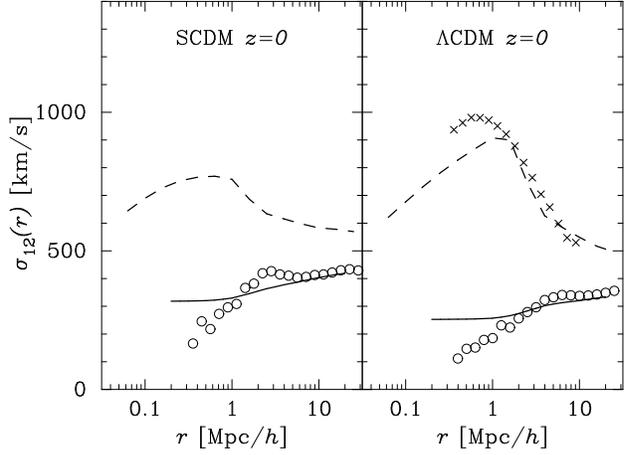,height=6cm,bbllx=68pt,bblly=59pt,bburx=608pt,bbury=452pt}}
\caption{Pairwise dispersions in the GIF simulations.  
Open circles and crosses show the streaming motions of the haloes 
and the dark matter, respectively.  Solid and dashed curves show 
predictions for the haloes and the dark matter (from Sheth et al. 2000).}
\label{sig12h}
\end{figure}
So far, we have shown how the mean streaming motions of the dark 
matter and the haloes are related.  Sheth et al. (2000) discuss how to 
do this for the second moment of the pairwise velocity distribution.  
They argued that the dark matter particles receive substantial nonlinear 
kicks to their initial velocities (essentially, the virial motions 
within haloes), whereas the haloes do not (Sheth \& Diaferio 2000).  
As a result the pairwise dispersion of the dark matter should be 
significantly larger than that of the haloes on all scales.  

Fig.~\ref{sig12h} compares what their equation~(31) predicts with the 
simulations (we refer the reader to their paper for details of the model).  
The open circles and crosses show the pairwise dispersion, 
$\sigma_{12}(r)$, of the haloes and the dark matter respectively, 
and the dashed and solid curves show the model predictions.  (We 
do not show $\sigma_{12}$ for the dark matter in the SCDM simulations 
because the simulation box is sufficiently small [85Mpc$/h$]
that cosmic variance affects the measurement significantly.)  The model 
is reasonably accurate on large scales, and not accurate on small scales.  
Sheth et al. (2000) discuss why this happens for the dark matter 
(the inaccuracy is due to the simplifing assumptions that haloes have 
no substructure, the pairwise dispersion from a single halo is isotropic 
and independent of the pair position, and the number of pairs in the 
infall regime around haloes is sufficiently small that the use of 
virial motions to model the dispersion from infalling pairs does not 
not lead to a large error).  
For the haloes, this discrepancy appears on scales which are of the order 
of a typical $m_*$ halo and smaller.  This suggests that the discrepancy 
almost surely arises from using linear theory to model the spatial 
distribution and velocities of haloes on scales which are smaller 
than the smoothing scale used to make the model prediction.  
Despite the quantitative discrepancies on small scales, the model is in 
qualitative agreement with the simulation:  the pairwise dispersion of 
the haloes is substantially smaller than that of the dark matter.

\section{Galaxies}\label{gals}
The previous section showed that the first moment of the pairwise velocity 
distribution of haloes is different from that of the dark matter.  
It showed that massive haloes separated by large distances are streaming 
together more rapidly than less massive haloes at the same separation, 
and that this difference scaled with one rather than two powers of the 
halo bias factor.  It also showed that the dark matter statistic was 
obtained by weighting the halo statistic by the number of dark matter 
particle pairs per halo.  The second moments of the pairwise velocity 
distributions are also different.  In this case, also, the dark matter 
statistic is got by weighting by the number of particle pairs per halo.  
However, the pairwise dispersion is also sensitive to the fact that virial 
motions within haloes can be substantially higher than the motions of the 
haloes themselves.  As a result, the pairwise velocity dispersion of dark 
matter particles is substantially larger than that of haloes, on all 
scales.  This section studies what these results imply for the pairwise 
motions of galaxies.  

We will model galaxies as random particles in dark matter haloes.  
That is, the motion of the galaxies is the same as that of the dark 
matter particle with which they are associated.  In this sense there 
is no velocity bias in our model; the fact that velocity statistics for 
the dark matter and the galaxies may, nevertheless, be different, arises 
solely from the fact that dark matter statistics weight each halo 
proportional to halo mass, whereas galaxy statistics do not.  
Such models for the difference between the statistics of galaxies 
and dark matter particles have received considerable attention 
recently.  Seljak (2000), Peacock \& Smith (2000) and 
Scoccimarro et al. (2000) have used them to model the spatial 
distribution of galaxies, Sheth \& Diaferio (2000) describe how to 
model the distribution function of galaxy peculiar velocites, and 
Sheth et al. (2000) describe how to use these models to do 
analytically what Jing, Mo \& B\"orner (1998) did numerically in 
their study of the pairwise velocity dispersion of galaxies.  

Within the context of this model, galaxies are treated by setting 
\begin{equation}
v_{12}^{\rm gal}(r) =  v_{12}^{\rm 1gal}(r) + v_{12}^{\rm 2gal}(r),
\label{v12gal}
\end{equation}
where the two terms denote the contribution to the statistic from 
galaxies in the same and in different haloes, respectively.  
The second term on the right hand side can be got by modifying 
equation~(\ref{v12twoh}):  
\begin{eqnarray}
{v_{12}^{\rm 2gal}(r)\over Hr} &=& 
\int\! {\rm d}m_2 \int\! {\rm d}m_1\,
{g(m_1) n(m_1)\over\bar\rho_{\rm gal}}
{g(m_2) n(m_2)\over\bar\rho_{\rm gal}}\nonumber\\
&&\ \times\ 
{[1 + b(m_1)b(m_2)\xi^{\rm Lin}_{\rm dm}(r)]\over 1 + \xi_{\rm gal}(r)}\,
{v_{12}^{\rm halo}(r)\over Hr} 
\label{v12g}
\end{eqnarray}
where
\begin{eqnarray}
\bar\rho_{\rm gal} &=& \int {\rm d}m\,g(m) n(m) \qquad
{\rm and} \nonumber\\
\xi_{\rm gal}(r) &=& \int {\rm d}m\,{g_2(m) n(m)\over\bar\rho_{\rm gal}}
{\lambda(r|m)\over\bar\rho_{\rm gal}}\nonumber \\
&&+\int {\rm d}m_2 \int {\rm d}m_1\,
{g(m_1) n(m_1)\over\bar\rho_{\rm gal}}
{g(m_2) n(m_2)\over\bar\rho_{\rm gal}}\nonumber\\
&&\ \times\ 
\Bigl[b(m_1)b(m_2)\xi^{\rm Lin}_{\rm dm}(r)\Bigr]\nonumber \\
&\equiv& \xi^{\rm 1halo}_{\rm gal}(r) + b_{\rm gal}^2\,\xi^{\rm Lin}_{\rm dm}(r).
\label{xig}
\end{eqnarray}
Here $g(m)$ and $g_2(m)$ denote the first and second moments 
of the distribution of the number of galaxies in $m$-haloes, and we 
set $g_2(m)=0$ if $g(m)<1$.  
There are details associated with how one treats the central galaxy 
in a halo, but, for the most part, these amount to a small effect 
(see Sheth \& Diaferio 2000), so we have ignored them---they add 
complications but no essential change to the logic of our argument.  

On scales larger than a few Mpc$/h$, $v^{\rm 2gal}_{12}$ dominates 
over the one-halo contribution.  
If we assume linear theory for the evolution of the two-halo term 
(equation~\ref{v12lin}), then we can set 
 $2f(\Omega)\bar\xi^{\rm Lin}_{\rm dm}/3 \to -v_{12}^{\rm dm}/Hr$ times 
$1+\xi^{\rm Lin}_{\rm dm}(r)$, and then equation~(\ref{v12g}) reduces to 
\begin{equation}
{v_{12}^{\rm 2gal}(r)\over Hr} \approx 
{v_{12}^{\rm dm}(r)\over Hr}\ b_{\rm gal}\ 
\left[{1+\xi^{\rm Lin}_{\rm dm}(r)\over 1 + \xi_{\rm gal}(r)}\right].
\label{v12twog}
\end{equation}
This shows that, on large scales, the streaming motions of galaxies 
can be biased relative to the dark matter.  The extent to which they 
are biased is related to how differently they are clustered, and this, 
in turn, depends on the $g(m)$ relation.  

On smaller scales, the streaming motions are dominated by galaxy 
pairs in which both members are in the same halo.  A little thought 
shows that this can be computed simply by setting 
$\bar\rho\to\bar\rho_{\rm gal}$ and $m^2\to g_2(m)$ in 
equation~(\ref{lamseq}).  
This is because the $g(m)$ relation does not introduce any additional 
time dependence---recall that the number density of haloes in the present 
model is fixed to the value it has today, and the galaxies are to be 
thought of simply as marked tracer particles within the haloes.  

This has an interesting consequence.  Suppose one wishes to use the 
pair conservation equation~(\ref{pairs}) to estimate $v_{12}^{\rm gal}(r)$.  
Then the model above suggests that, on small scales, simply inserting the 
observed galaxy correlation function into equation~(\ref{pairs}) should 
be reasonably accurate.  However, on larger scales, doing this leads 
to an estimate of $v_{12}^{\rm gal}$ which is incorrect, for the 
following reason.  Because there is no time dependence in $g(m)$, 
the result of doing this has the same time dependence as in the 
case of the dark matter for which $g(m)=m$, or the case of the haloes 
for which $g(m)=1$.  But we know that, for the haloes, doing this results 
in the wrong answer, because it neglects the evolution of the bias factor.  
The case of galaxies is no different; if we insert the observed galaxy 
correlation function into the pair conservation equation, then we are 
incorrectly neglecting the evolution of the bias factor of the galaxies.  
This would lead one to conclude, incorrectly, that $v_{12}$ should scale 
as $b_{\rm gal}^2$ rather than as $b_{\rm gal}$.  

\begin{figure}
\centering
\mbox{\psfig{figure=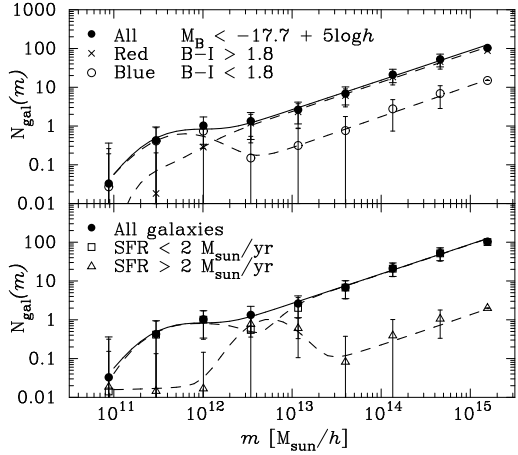,height=6cm,bbllx=87pt,bblly=56pt,bburx=608pt,bbury=515pt}}
\caption{Mean number of bright galaxies as a function of parent halo 
mass in the $\Lambda$CDM GIF semianalytic galaxy formation model of 
Kauffmann et al. (1999).  Top panel shows the result of dividing the 
sample into two based on colour.  Bottom panel shows a division based 
on star formation rate.  Crosses, circles, squares and triangles are 
for objects classified as being red, blue, quiescent and star-forming 
galaxies respectively.  }
\label{gifngmh}
\end{figure}

\begin{figure}
\centering
\mbox{\psfig{figure=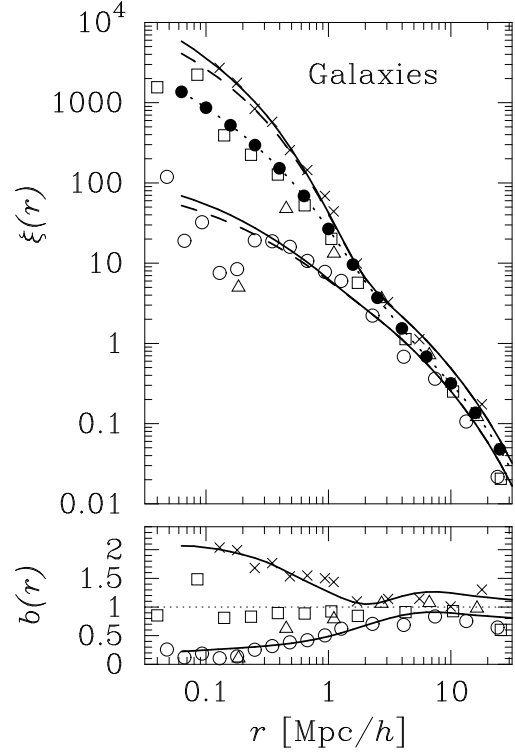,height=10cm,bbllx=69pt,bblly=59pt,bburx=386pt,bbury=521pt}}
\caption{Correlation functions of different tracers of the dark matter 
density field in the $\Lambda$CDM GIF semianalytic galaxy formation model.  
Filled circles are for the dark matter, crosses are for red galaxies,
squares for galaxies which have low star formation rates, 
triangles for galaxies with high star formation rates, 
and open circles for blue galaxies.  The two solid curves show our 
model predictions for the red and blue galaxies, and the dashed curves 
show what happens if we use the second factorial moment of the galaxy 
counts, rather than the second moment when making our model predicition.  
For comparison, the dotted curve shows the predicted dark matter 
correlation function.  }
\label{xigals}
\end{figure}

\subsection{Comparison with simulations}\label{sims}
To illustrate how our model works, we will use the $g(m)$ relations 
we obtained from the semianalytic GIF $\Lambda$CDM model of 
Kauffmann et al. (1999) which are now publically available.  
Sheth \& Diaferio (2000) provide a fitting formula for the $g(m)$ 
relation of a GIF galaxy catalog which was constructed by choosing 
all galaxies brighter than $M_{\rm V} = -17.7 + 5\log h$, after 
accounting for the effects of dust.  We divided that catalog up 
into two subsamples based on colour (galaxies labelled as being 
redder or bluer than $B-I=1.8$) and on star-formation rate 
(rates greater or less than $2 M_\odot/$yr).  
Fig.~\ref{gifngmh} shows these relations.  The dashed lines show 
the following fits:
\begin{eqnarray}
N_{\rm All}(m) &=& (m_{11}/700)^{0.9} 
             + 0.5\,{\rm e}^{-4[\log_{10}(m_{11}/5.6)]^2}\nonumber \\
          && + (m_{11}/30)^{0.75}{\rm e}^{-(2/m_{11})^2} \nonumber\\
N_{\rm Blue}(m) &=& (m_{11}/500)^{0.8} 
             + 0.6\,{\rm e}^{-4[\log_{10}(m_{11}/6.2)]^2}\nonumber \\
N_{\rm Red}(m) &=& N_{\rm All}(m) - N_{\rm Blue}(m) \nonumber \\
N_{\rm hSFR}(m) &=& 0.015 + (m_{11}/7000)^{0.9} 
             + {\rm e}^{-8[\log_{10}(m_{11}/56.2)]^2}\nonumber\\
N_{\rm lSFR}(m) &=& N_{\rm All}(m) - N_{\rm hSFR}(m)
\end{eqnarray}
where $m_{11}$ is the halo mass in units of $10^{11}M_\odot/h$.  
The solid lines are the same in both panels; they show $N_{\rm All}(m)$, 
and they equal the sum of the two dashed lines.  These relations 
can be used to compute the statistics of one of these galaxy 
samples, rather than dark matter particles, by setting $g(m)$ equal to 
the appropriate $N_{\rm gal}(m)$ relation.  

In addition to these mean $N_{\rm gal}(m)$ relations, our models also 
require the second moment of the number of galaxies per halo 
distribution.  We have approximated it by setting $g_2(m)=0$ if 
$g(m)<1$, and 
\begin{equation}
g_2(m) = \mu^2(m)\,g^2(m) + \mu\,g(m)\qquad{\rm if} g(m)\ge 1,
\end{equation}
where $\mu(m) = \log_{10}[20\,g(m)]^{1/2}$ if $g(m)\le 5$, and $\mu=1$ 
when $g(m)$ is larger.  Because $\mu=1$ for a Poisson distribution, this 
approximately accounts for the fact that the scatter in galaxy counts is 
sub-Poisson in low mass haloes.  If we use the same galaxy sample that 
Scoccimarro et al. (2000) did, then our model for the scatter is similar 
to theirs.  

\begin{figure*}
\centering
\mbox{\psfig{figure=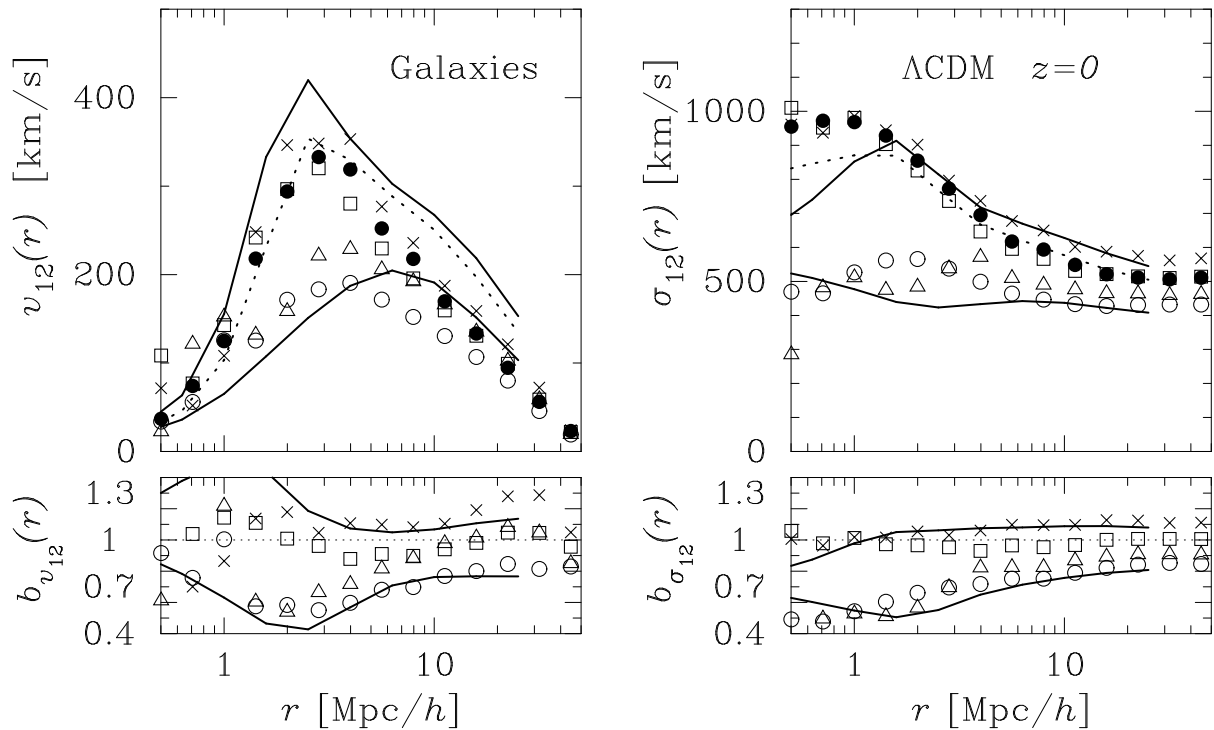,height=10cm,bbllx=69pt,bblly=58pt,bburx=721pt,bbury=465pt}}
\caption{The mean streaming velocity (left) and pairwise velocity 
dispersion (right) in the $\Lambda$CDM GIF semianalytic galaxy 
formation model.  Filled circles are for the dark matter, 
crosses are for red galaxies,
squares for galaxies which have low star formation rates, 
triangles for galaxies with high star formation rates, 
and open circles for blue galaxies.  The solid and dotted curves 
show our predictions for the red and blue galaxies, and the 
dark matter, respectively.  }
\label{gifgals}
\end{figure*}

The correlation functions for these four subsamples are shown in 
Fig.~\ref{xigals}:  crosses, open circles, triangles and squares 
show $\xi_{\rm gal}(r)$ for red, blue, star-forming and quiescent 
galaxies in the GIF simulation.  Filled circles show the correlation 
function of the dark matter particles.  
Notice how similar the correlation function of the blue sample is 
to that of the star-forming sample, how similar the red and quiescent 
samples are, and how different the blue and star-forming samples are 
from the red and quiescent samples.  

The two solid curves show the result of using our model to compute the 
correlation functions of the red and blue samples, because these differ 
the most from each other.  We did this by setting $g(m)$ in 
equation~(\ref{xig}) equal to the appropriate $N_{\rm gal}(m)$ relation.  
For comparison, the dashed curves show the result of using the second 
factorial moment, rather than the second moment, when computing the galaxy 
correlation functions; the difference only matters on small scales.  
The dotted curve shows our calculation of $\xi_{\rm dm}(r)$ which has 
$g(m)=m$.  The bottom panel shows how 
 $b\equiv \sqrt{\xi_{\rm gal}/\xi_{\rm dm}}$ 
(the ratio of the solid and dotted curves) depends on scale.  
Both panels show that our model provides a good description of the 
simulation results.  In computing our model predictions, 
we assumed that the two samples both trace their parent dark matter haloes 
similarly.  That is, we used the same function $\lambda(r|m)$ for both 
the red and the blue samples.  If the red galaxies were more centrally 
concentrated than the blue, we could have incorporated it into our 
analysis by adjusting $\lambda(r|m)$.  In fact, Fig.~2 of 
Diaferio et al. (1999) shows that, in the semianalytic model, the red 
galaxies in massive clusters are concentrated more towards the centres 
of their parent halos than the blue ones are.  The agreement between the 
simulation and our model curves in which we made no such adjustment for 
this suggests that it must amount to only a weak effect.  

We turn, therefore, to the first and second moments of the pairwise 
velocity distribution for these four subsamples.  
Fig.~\ref{gifgals} shows results for the same semianalytic galaxy 
samples shown in Fig.~\ref{xigals}.  As before, crosses, squares, 
triangles and open circles show galaxies classified as being red, 
quiescent, star-forming, or blue.  Filled circles show the corresponding 
statistics of the dark matter particles.  
As for the correlation functions, the blue and star-forming samples 
are quite different from the red and quiescent samples.  Our model 
shows that this arises simply from the fact that these samples have 
rather different $N_{\rm gal}(m)$ relations---there are only a few
blue, star forming galaxies in clusters.

Notice that blue galaxies (circles) have the smallest streaming motions, 
and red galaxies (crosses) have the largest $v_{12}$ values. 
Our model predicts that larger streaming motions at large separations 
indicate a higher amplitude of clustering on those scales.  Comparison 
with Fig.~\ref{xigals} shows that the correlation functions of the red 
galaxies are biased high relative to the dark matter, whereas the blue 
galaxies are biased low.  This is in qualitative agreement with our 
model.  

The solid lines in Fig.~\ref{gifgals} provide a more quantitative 
comparison between our model predictions for the red and blue galaxies, 
and the values of the galaxy velocities measured in the GIF simulation.  
The model predictions are in reasonable agreement with the simulation, 
although the agreement is certainly not as good as it was for the 
correlation functions.  Sheth et al. (2000) discuss the reason for the 
overestimate in $v_{12}(r)$ on large scales (e.g., these models do not 
satisfy the integral constraint).  The bottom panels show the ratio 
of the galaxy velocities to those of the dark matter; i.e., 
$b_{v_{12}}\equiv v_{12}^{\rm gal}/v_{12}^{\rm dm}$, and similarly 
for $b_{\sigma_{12}}$.  This ratio is scale dependent on smaller scales.  
Our model describes the scale dependence reasonably well.  

\section{Discussion}\label{discuss}
We presented a simple model of how the streaming motions of haloes 
and galaxies depends on separation.  We tested the model using 
the publically available simulations of Kauffmann et al. (1999).  
In the semi-analytic galaxy formation model, the mean streaming 
motions depend rather strongly on how the galaxy sample was selected.  
For example, blue galaxies have smaller streaming motions than red 
galaxies.  We showed that our model was able to describe the differences 
between a wide range of simulated galaxy catalogues rather well 
(Fig.~\ref{gifgals}).  

Our model predicts a very close relationship between the 
streaming motions of the galaxies and their spatial distribution.  
Optical and IRAS galaxies cluster differently 
(e.g. Marzke et al. 1995; Fisher et al. 1994).  Therefore, they 
must be biased differently relative to the dark matter.  
If the streaming motions of optical galaxies are the same as 
the dark matter, then our model predicts that the streaming motions 
of IRAS galaxies must be different from that of the dark matter.  In 
other words, whether or not the streaming motions of a given galaxy 
sample trace the motions of the underlying dark matter depends on how 
the sample was selected.  Thus, in the absence of strong arguments for 
why a given galaxy sample is expected to be unbiased, one should be 
cautious when interpretting measurements of the streaming motions of 
galaxies.  

If the correlation function and the streaming motions of two different 
galaxy samples have been measured, then the model described here 
(equations~\ref{v12gal} and~\ref{v12twog}) says that the square 
root of the ratio of the two correlation functions 
(at, say, 20 Mpc/$h$) should equal the ratio of the streaming motions 
on the same scale.  This can be used to test the validity of the model.  
Again, however, caution is required because this relationship is only 
true on large scales.  For the semi-analytic galaxy samples we presented, 
this simple linear biasing was a good approximation on scales larger than 
about 10 Mpc$/h$ (although our model is able to describe the scale 
dependence of this ratio even on smaller scales).  In this context, we 
think it worth noting that the semi-analytic galaxy samples we presented 
here cannot explain the results of Juszkiewicz et al. (2000) who found that 
ellipticals and spirals have the same value of $v_{12}$ on separations of 
about 10 Mpc/$h$, even though they estimate that the spatial distributions 
have bias factors which differ by a factor of two:  $b_E/b_S\approx 2$.  
It will be interesting if future data sets confirm this.  

In addition to studying how the mean streaming velocity depends on 
galaxy type, our Fig.~\ref{gifgals} also shows that the second moment 
of the pairwise velocity distribution depends strongly on galaxy type.  
For example, our model of the pairwise dispersion suggests that the 
dispersion of blue galaxies should be substantially smaller than 
that of red ones (although this difference depends on the colour cut), 
especially on scales of 1 Mpc$/h$ or so.  Thus, it is not surprising 
that $\sigma_{12}$ for optical galaxies (Marzke et al. 1995) is almost a 
factor of two larger than for IRAS galaxies (Fisher et al. 1994).  
The effects of redshift space distortions are larger if the pairwise 
dispersion is larger.  This means that, on small scales, the amplitude 
of the redshift space correlation function of red galaxies should
be substantially smaller than the real space correlation function, 
but the difference between real and redshift space correlation functions 
of blue galaxies should not be as dramatic.  
This is a generic prediction of these sorts of galaxy formation models.
This suggests that galaxy redshift samples cut by colour should 
provide a useful and direct test of these models.  

Dividing a galaxy sample by colour allows another simple test of 
these models.  At large separations, where most pairs are in separate 
haloes, the model described above predicts that the cross-correlation 
function of the two colour samples should simply be the geometric mean 
of the two individual samples.  
If the two galaxy samples both trace their parent dark matter haloes 
in the same way (our Fig.~\ref{xigals} shows that this assumption describes 
the semianalytic model well), then this will be approximately true even 
on smaller scales.  (It will not be exactly true because the scatter 
in the $N_{\rm gal}(m)$ relations are, typically, sub-Poisson.)  
Data sets currently available should be able to test this prediction.  

\section*{Acknowledgments}
This collaboration was started at the German American Young Scholars
Institute in Astroparticle Physics, at the Aspen Center for Physics
in the fall of 1998, and at Ringberg Castle in 1999.
RKS thanks the IAS, as well as the University and Observatory at Torino, 
and LH \& RS thank Fermilab for hospitality where parts of this work 
were done.  We all thank the Halo Pub for inspiring nourishments.
The N-body simulations, halo and galaxy catalogues used in this paper
are publically available at {\tt http://www.mpa-garching.mpg.de/NumCos}.
The simulations were carried out at the Computer Center of the 
Max-Planck Society in  Garching and at the EPCC in Edinburgh, as part 
of the Virgo Consortium project.  In addition, we would like to thank 
the Max-Planck Institut f\"ur Astrophysik where some of the computing 
for this work was done.  
RKS is supported by the DOE and NASA grant NAG 5-7092 at Fermilab. 
LH is supported by NASA grant NAG5-7047, NSF grant PHY-9513835
and the Taplin Fellowship. He also thanks the DOE for an 
Outstanding Junior Investigator Award (grant DE-FG02-92-ER40699)
RS is supported by endowment funds from the IAS and by NSF grant 
PHY-0070928 at IAS.

\appendix
\section{The evolution of the correlation function}
This Appendix shows how the correlation function and the mean 
streaming motions evolve in our model.  

The correlation function on comoving scale $x$ when the expansion 
factor is $a$ is the sum of two terms:  
\begin{equation}
\xi_{\rm dm}(x,a) = \xi_{\rm dm}^{\rm 1halo}(x,a) + 
                    \xi_{\rm dm}^{\rm 2halo}(x,a),
\end{equation}
where 
\begin{equation}
\xi_{\rm dm}^{\rm 2halo}(x,a) = 
\xi_{\rm dm}^{\rm Lin}(x,a)\,\left[\int {\rm d}m\,f(m)\,b(m,a)\right]^2
\label{xidm2h}
\end{equation}
(compare equation~\ref{xig}), and 
\begin{equation}
\xi_{\rm dm}^{\rm 1halo}(x) = \int {\rm d}m\, {m^2n(m)\over\bar\rho} \,
{\lambda(x|m)\over\bar\rho}.
\end{equation}
For haloes of virial mass $m$ which have density profiles 
of the form given by Navarro , Frenk \& White (1997), 
\begin{equation}
\lambda(x|m) = {c^3\,\phi^2(c)\over 4\pi r_{\rm vir}^3}\,
g\left({cx\over x_{\rm vir}}\right) ,
\end{equation}
where $g(y)$ is given in Appendix~A3 of Sheth et al. (2000).  
Here $r_{\rm vir}$ is the virial radius of 
the halo in proper physical coordinates, $x_{\rm vir}=r_{\rm vir}/a$ 
is the comoving virial radius, and $c(m)$ is a parameter which describes 
how centrally concentrated the halo profile is: it is the ratio of the 
virial radius to a central core radius, and it depends on the halo mass.
The normalization term is 
\begin{equation}
\phi(c) \equiv \Bigl[\ln (1+c) - c/(1+c)\Bigr]^{-1}.
\end{equation}
The mean streaming motions from pairs in the same halo is given by 
equation~(\ref{lamseq}) in the main text.  

We would like to study what happens if we make the profile of 
the region containing $m$ depend on time in the following way:  
we would like to use the same family of profiles, such as the NFW 
set, to describe the density run at any time, and we want to parametrize 
the evolution of the profile shape by changing the values of the 
profile's parameters.  There is no physical reason why the NFW 
form should describe the density run around a region which has yet 
to virialize; we are only using this to illustrate how our argument works.  

For NFW profiles, this means that we will think of $X(a)=X_a$ as 
the comoving boundary of the region containing $m$ at the time when 
the expansion factor is $a$:  
initially $X(a_i) = X_i$, and today $X(a_0) = x_{\rm vir}$.  
In addition, we will allow the concentration parameter to depend on 
mass as well as time:  $c(m,a)=c_a$.  The idea is that the density run 
around the centre of the region of radius $X_a$ containing $m$ was 
presumably differently concentrated at $a_i$ than it is today.  
In particular, we would like to see if this model can produce the 
same streaming motions as equation~(\ref{lamseq}) in the main text.  

To show what is required for this to happen, set $mn(m)/\bar\rho = f(m)$, 
and also set $m = 4\pi X_i^3\bar\rho_0/3$, where $X_i$ is the initial 
comoving radius of the halo, and $\bar\rho_0=\bar\rho/a^3$ is the 
comoving density of the background.  Then the expression for the 
correlation function becomes 
\begin{equation}
\xi^{\rm 1halo}_{\rm dm}(x) = 
\int {\rm d}m\,f(m)\,\left({c_a\,X_i\over X_a}\right)^3
\,{\phi^2(c_a)\over 3}\,g\left({c_a x\over X_a}\right) .
\label{xixa}
\end{equation}
The volume integral of this is 
\begin{eqnarray}
\bar\xi (x) &=& \int {\rm d}m\,f(m)\,\left({c_a\,X_i\over X_a}\right)^3
\,{\phi^2(c_a)\over x^3}\int_0^x {\rm d}y \,y^2\,
g\left({c_a y\over X_a}\right) 
\nonumber \\
&=& \int {\rm d}m\,f(m)\,\phi^2(c_a)\,\left({X_i\over x}\right)^3
\bar g\left({c_a x\over X_a}\right),
\end{eqnarray}
where we have defined $\bar g(y)\equiv\int_0^y {\rm d}z\,z^2 g(z)$.  
So 
\begin{eqnarray}
{\partial \bar\xi (x)\over\partial\ln a} &=& 
\int\!{\rm d}m\,f(m)\,\Biggl[{\partial \ln \phi^2(c_a)\over\partial\ln a}
\left({X_i\over x}\right)^3\phi^2(c_a)\,
\bar g\left({c_ax\over X_a}\right)\nonumber\\ 
&& + {\partial \ln (c_a/X_a)^3\over\partial\ln a}{\phi^2(c_a)\over 3}
\left({c_aX_i\over X_a}\right)^3
g\left({c_ax\over X_a}\right)\Biggr].
\end{eqnarray}
This will result in the same streaming motions as 
equation~(\ref{lamseq}) if 
\begin{eqnarray}
{\partial {\,\rm ln}\phi^2(c_a) \over \partial {\,\rm ln} a}
&=& {6f_\Omega\over 3+n_*}\qquad {\rm and}\nonumber\\
{\partial {\,\rm ln}(c_a/X_a)^3 \over \partial {\,\rm ln} a}
&=& - {6f_\Omega\over 3+n_*},
\label{constraints}
\end{eqnarray}
where $n_*$ is the slope of the linear power spectrum on the scale 
on which the rms density fluctuation is 1.686, and 
$f_\Omega\equiv \partial\ln D(a)/\partial\ln a \approx \Omega^{0.6}$, 
where $D(a)$ is the linear theory growth factor.  
This sums up what is required of the time dependence of the 
concentration $c_a$ and the comoving radius $X_a$ containing $m$.
Note that these requirements are non-trivial, because
both $c$ and $X$ depend on the mass $m$, but we are requiring 
that the derivatives work out to be independent of $m$.  

We can see what this implies for the evolution of the correlation 
function if we insert the scalings of equation~(\ref{constraints}) in 
equation~(\ref{xixa}):
\begin{displaymath}
\xi_{\rm dm}^{\rm 1halo}(x,a)\! =\! 
\int {\rm d}m\,f(m)\left({c_0\,X_i\over X_0}\right)^3
{\phi^2(c_0)\over 3}\,g\left({c_0 x/X_0\over D^{2/(3+n)}}\right) ,
\label{xix0}
\end{displaymath}
where the subscript `0' denotes the values of quantities at the 
present time.  
This shows that the number of pairs on comoving scale $x$ at the 
present time is the same as the number of pairs which, at the earlier 
time when the linear growth factor was $D=D(a)/D_0$ were on the larger 
comoving scale $x/D^{2/(3+n)}$.  

The correlation function decreases monotonically with scale, 
so the expression above implies that $\xi^{\rm 1halo}_{\rm dm}$ 
was smaller at early times than it is today.  At very early times,
therefore, the correlation function might plausibly be dominated 
by the two-halo term (equation~\ref{xidm2h}).  The evolution of this 
term can be got from inserting the evolution of the bias factor 
(equation~\ref{biasm}) into it.  The integrals over $m$ can be done 
analytically, with the result that 
$\xi_{\rm dm}^{\rm 2halo}(x,a) = D^2\, \xi_{\rm dm}^{\rm Lin}(x,a_0)$.
At sufficiently early times, this two-halo term dominates on all 
scales, so our model for the correlation function reduces 
correctly to the linear theory expression.  

Our requirement that the same NFW form hold at all times means that 
the profile shape evolves as 
\begin{eqnarray}
{\rho_a(s)\over\bar\rho} 
&=& {\phi(c_a)/3\over s(s + (X_a/c_a)/X_i)^2}\nonumber \\
&=& {\rho_{a_0}(S)\over\bar\rho_0} D^{3/(3+n_*)}
    \left({S + 1/c_0\over S + D^{2/(3+n)}/c_0}\right)^2,
\end{eqnarray}
where $s$ is the comoving distance from the centre in units of the 
initial comoving scale $X_i$, 
whereas $S$ is in units of the virial radius at $a_0$.  
The second equality follows from the scalings above for the evolution 
of $\phi(c_a)$ and $X_a/c_a$, and setting $D\equiv D(a)/D_0$.  
The profile evolves in such a way that the density on scales 
$s>(X_a/c_a)/X_i$ grows as $a$ increases, as we expect.  On much smaller 
scales, $s\ll (X_a/c_a)/X_i$, and the profile shape is more like 
$(c_aX_i/X_a)^2\,\phi(c_a)/S\propto D^{-1/(3+n)}/S$:
on very small scales, the density decreases with time!  

\begin{figure}
\centering
\mbox{\psfig{figure=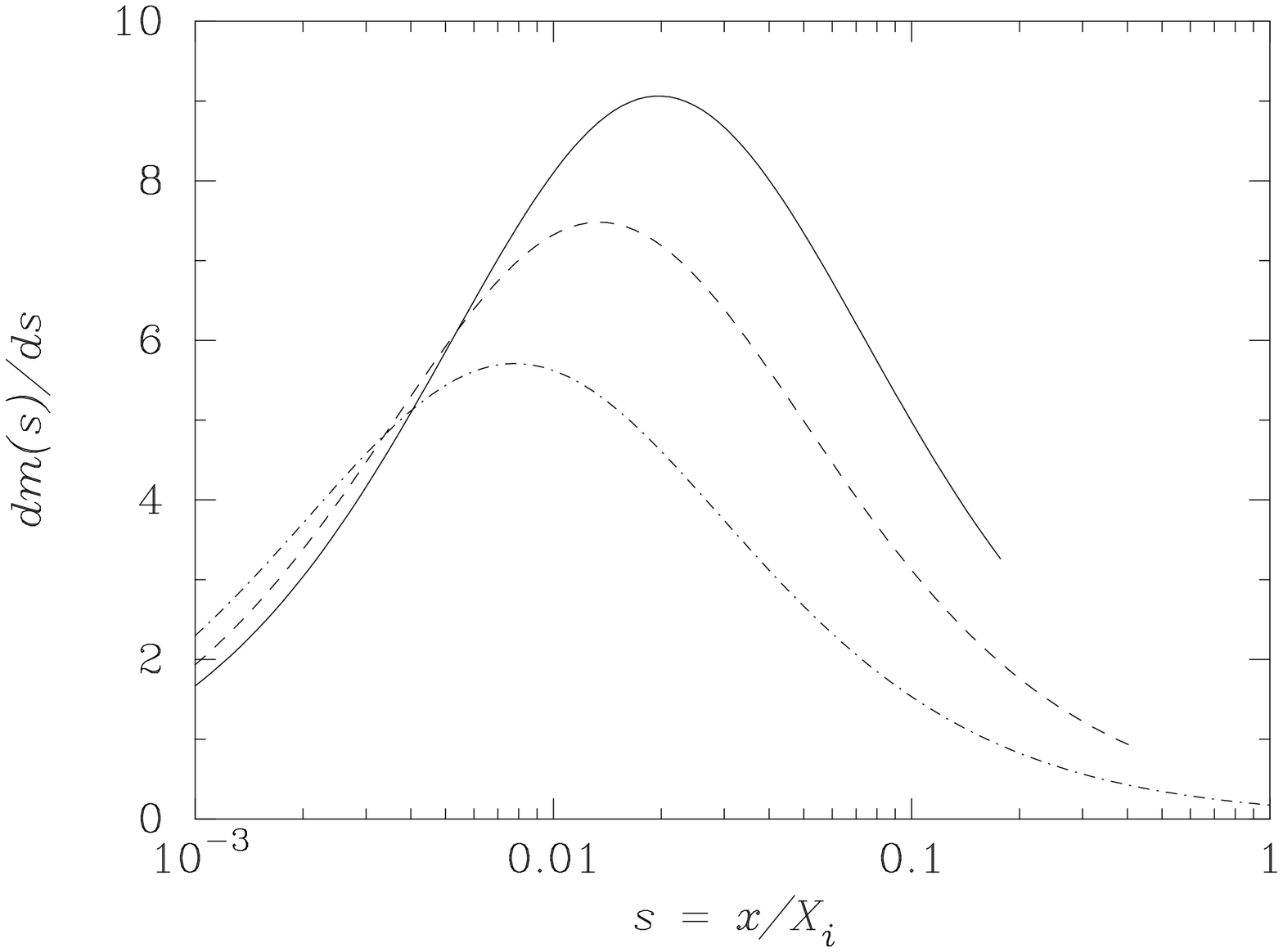,height=6cm,bbllx=69pt,bblly=56pt,bburx=611pt,bbury=459pt}}
\caption{The evolution of the mass profile if we require that the 
density profile have the NFW form at all times.  For most of the 
comoving volume, the mass in a given comoving shell increases with time.  
Only within the core of the object does the mass decrease with time.  
We have truncated the profiles at the point where they all enclose 
the same mass $m$.}
\label{nfwev}
\end{figure}

Because this small scale behaviour of the density seems contrary to our 
intuition, we thought it worth studying the evolution of the mass as a 
function of comoving scale.  The scalings above imply that the fraction 
of the total mass $m=M(s)/M$ which is in the range d$s$ around $s$ from 
the halo centre when the growth factor is $D\equiv D(a)/D_0$ is 
\begin{equation}
{{\rm d}m(s)\over {\rm d}s} \propto 
{s \phi(c_0) D^{3/(3+n_*)}\over 
\Bigl[s + D^{2/(3+n_*)}/[c_0(\Delta_{\rm nl}/\Omega)^{1/3}]\Bigr]^2}.
\label{dmds}
\end{equation}
Fig.~\ref{nfwev} shows an example of how the mass gets redistributed 
as the profile evolves.  It was constructed by setting $n_*=-1.5$, $c_0=9$ 
and $D_{\rm nl}/\Omega=180$ in equation~(\ref{dmds}) (the first two values 
approximate those of an $m_*$ halo in a $\Lambda$CDM simulation).  
The solid, dashed, and dot-dashed curves show equation~(\ref{dmds}) at 
$D=1$, $0.75$ and $0.5$ respectively.   
For most of the volume of the halo, the mass in a given comoving 
shell increases as $D$ increases.  Only well within the core of the 
object does it decrease with time.  We only show the shape of the profile 
out to the radius $X_a$ which contains the mass $m$.  At $D=1$, $X_a$ is 
the virial radius, which is at $s = 1/5.6\approx 0.18$; $X_a$ was larger 
earlier, so that the total mass contained in the profile remains constant.  
By $D=0.6$ or so $X_a>1$, indicating that the model profile must extend 
beyond $X_i$ if it is to enclose mass $m$; at this point the model has 
really broken down.  

A similar analysis of the Hernquist profile shows the same qualitative 
features:
\begin{equation}
{\rho_a(S)\over\bar\rho} = {\rho_{a_0}(S)\over\bar\rho_0}D^{5/(3+n_*)}
\left(S + b_0\over S + b_0D^{2/(3+n_*)}\right)^3. 
\end{equation}
Presumably, this apparently unphysical behaviour is a consequence of 
our unphysical requirement that the profile have the same functional 
form at all times.

\end{document}